\documentclass[journal]{IEEEtran}

\usepackage{graphicx,cite,epsfig}
\usepackage{amsmath} % assumes amsmath package installed
\usepackage{amssymb}  % assumes amsmath package installed
\DeclareMathOperator *{\argmin}{argmin}
\usepackage{pifont}
\usepackage{stmaryrd}
\usepackage{bbding}
\usepackage{subfigure}
\usepackage{algorithm}
\usepackage{algorithmic}
\usepackage{array}
\usepackage{amsthm}
\usepackage{multirow}
\usepackage{url}
\usepackage{mathrsfs}
\usepackage{color, soul} %ÓÃcolor, ºÍ soul °ü
\usepackage{mathrsfs}
\usepackage{float}

\newcommand{\tabincell}[2]{\begin{tabular}{@{}#1@{}}#2\end{tabular}}

\usepackage{booktabs}
\usepackage{threeparttable}
\usepackage{amssymb}

\begin{document}
\title{Toward Terabits-per-second Communications: A High-Throughput Hardware Implementation of $G_N$-Coset Codes}

\author{\IEEEauthorblockN{Jiajie Tong, Xianbin Wang, Qifan Zhang, Huazi Zhang, Shengchen Dai, Rong Li, Jun Wang}

\IEEEauthorblockA{Huawei Technologies Co. Ltd.\\
Email: \{tongjiajie, wangxianbin1, Qifan.Zhang, zhanghuazi, daishengchen, lirongone.li, justin.wangjun\}@huawei.com}}

\maketitle
\thispagestyle{empty}

\begin{abstract}
Recently, a \emph{parallel} decoding algorithm of $G_N$-coset codes was proposed.
The algorithm exploits two equivalent decoding graphs.
For each graph, the inner code part, which consists of independent component codes, is decoded in parallel. The extrinsic information of the code bits is obtained and iteratively exchanged between the graphs until convergence.
This algorithm enjoys a higher decoding parallelism than the previous successive cancellation algorithms, due to the avoidance of serial outer code processing.
In this work, we present a hardware implementation of the parallel decoding algorithm, it can support maximum $N=16384$. We complete the decoder's physical layout in TSMC $16nm$ process and the size is $999.936\mu m\times 999.936\mu m, \,\approx 1.00mm^2$. The decoder's area efficiency and power consumption are evaluated for the cases of $N=16384,K=13225$ and $N=16384, K=14161$. Scaled to $7nm$ process, the decoder's throughput is higher than $477Gbps/mm^2$ and $533Gbps/mm^2$ with five iterations.
\end{abstract}

\section{Introduction}\label{section_introductions}
\subsection{Background}
High throughput is one of the primary targets for the evolution of mobile communications.
The next generation of mobile communication, i.e., 6G, is expected to supply $1$ Tbp/s throughput\cite{6G}.
which requires roughly a $100$x increase in throughput over the 5G standards.

$G_N$-coset codes, defined by Ar{\i}kan in \cite{ArikanPolar}, are a class of linear block codes with the generator matrix $G_N$.
$G_N$ is an $N \times N$ binary matrix defined as $G_N \triangleq F^{\otimes n}$,
in which $N=2^n$ and $F^{\otimes n}$ denotes the $n$-th Kronecker power of $F=[\begin{smallmatrix}
1 & 0  \\
1 & 1
\end{smallmatrix} ]$.

Polar codes is a specific type of $G_N$-coset codes, adopted for the 5G control channel, respectively.
The throughput of polar codes is limited by the successive cancellation (SC) decoders, since they are serial in nature.

\begin{figure}[]
	\centering
	\includegraphics[width=3.2in]{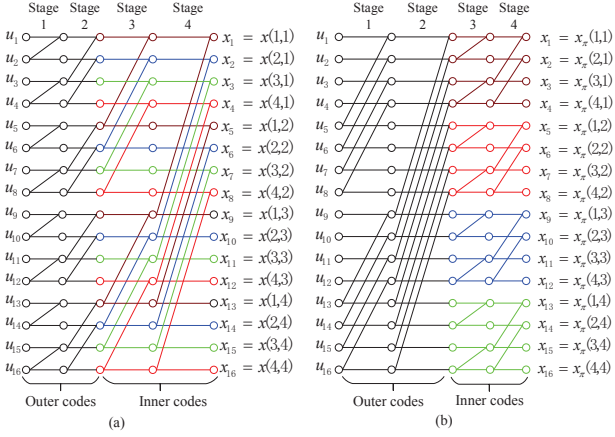}
	\caption{For $G_N$-coset codes, equivalent encoding graphs may be obtained based on stage permutations: (a) Ar{\i}kan's original encoding graph \cite{ArikanPolar} and (b) stage-permuted encoding graph.  Each node adds (mod-2) the signals on all incoming edges from the left and sends the result out on all edges to the right.  }
	\label{permutation}
\end{figure}

Recently, a parallel decoding framework of $G_N$-coset codes is proposed in \cite{TPC}. It is alternately decoded on two factor graphs $\cal{G}$ and $\cal{G}_{\pi}$, as shown in Fig.~\ref{permutation}. The permuted graph is generated by swapping the inner codes and outer codes. The decoder only decodes the inner codes of each graph $\Lambda \in \{\cal{G},\cal{G}_{\pi}\}$. In each $\Lambda$, the inner codes are $\sqrt{N}$ independent sub-codes that can be decoded in parallel. The code construction under the parallel decoding algorithm is different from polar/RM codes, and is studied separately in \cite{TPC}.

\subsection{Motivations and Contributions}
This paper introduce an ASIC implementation based on the parallel decoding framework (PDF). We set up a decoder which can support $N=16384$ $G_N$-coset codes. It deploys $\sqrt{N}=128$ sub-decoder to decode the 128 independent sub-codes in parallel. The target of high throughput and area efficiency is decomposed into the reduction of sub-codes decoding latency, worst-case iteration time, and chip area, and optimized respectively.

We adopt the proposal in \cite{SC_COMP} which employs successive cancellation (SC) decoders as the component decoder. It can support soft-in-hard-out decoding which results in low decoding complexity and reduced interconnection among the component decoders. In this work, we propose hardware-oriented optimizations on LLR generation and quantization. We implemented the whole decoder in hardware and present the ASIC layout to evaluate multiple key metrics. The hardware-specific data is obtained from the cells flip ratio from the circuit simulation results and the parasitic capacitance extracted from the layout result. With $16nm$ process, the area efficiency is $120Gbps/mm^{2}$, the power consumption is $100mW$ and energy efficiency is around $1pJ/bit$. Scaled to $7nm$, the area efficiency can reach $533Gbps/mm^{2}$ with five iterations.

\section{Parallel Decoding}
A parallel decoding framework is introduced in \cite{TPC}, where three types of component decoders (i) soft-output SC list (SCL), (ii) soft-output SC permutation list and (iii) soft cancellation (SCAN) are employed to decode the sub-codes (inner codes). To achieve even higher area efficiency, this work adopts SC~\cite{SC_COMP}, i.e., \emph{without list decoding}, as the sub-decoder.
In this section, we describe the parallel decoding framework-successive cancellation (PDF-SC) algorithm from the implementation point of view.

\label{section_decoding}
\subsection{Parallel decoding framework (PDF)}\label{sec:parallel_decoder}
We use $(i,j)$ to denote the $j$-th code bit position of $i$-th inner sub-code, and $k$ to denote the code bit position in the $G_N$-coset code. They have the following relationship
\begin{equation}\label{k_location}
	k=\left\{
	\begin{array}{lcl}
    j\times \sqrt{N} + i, \text{for}\;\Lambda=\cal{G},\\
	i\times \sqrt{N} + j, \text{for}\;\Lambda=\cal{G}_{\pi}.\\
	\end{array} \right.
\end{equation}

The aforementioned parallel turbo-like decoding framework is described in Algorithm~\ref{alg:stage-permute-decoder}.
In every two iterations, the algorithm \emph{alternately} decodes the two graphs $\cal G$ and ${\cal G}_\pi$ (line 4 in Algorithm~\ref{alg:stage-permute-decoder}) with $\sqrt{N}$ inner component decoders.
The $i$-th component decoder, denoted by $SubDecoder()$, is a SC decoder assisted by the error detector.

Note that each component decoder takes a soft input vector $L_{i}^t$, but outputs hard code bit estimates $\hat{c}$ and error detecting results $e$.
The mismatch between soft input and hard output poses a challenge for iterative decoding, since the SC component decoders in the next iteration cannot directly take the hard output from the previous iteration as input.
To solve this problem, \cite{SC_COMP} proposes to generate soft values from the hard outputs.
Specifically, the log likelihood ratio (LLR) of the $(i,j)$-th code bit in the $t$-th iteration, denoted by $L_{(i,j)}^t$, is calculated from the hard decoder outputs from the alternate graph (line 5).

For the $i$-th component decoder, the hard output vector and soft input vector have length $\sqrt{N}$
\begin{align*}
\hat{c}_{i}^t & \triangleq\{\hat{c}_{(i,j)}^t,j=0\cdots \sqrt{N}-1\}, \\
L_{i}^t & \triangleq\{L_{(i,j)}^t, j=0\cdots\sqrt{N}-1\},
\end{align*}
and the error detection indicator $e_i^t$ is a binary value.

The $\sqrt{N}$ independent inner sub-codes allow us to instantiate $\sqrt{N}$ component decoders for maximum degree of parallelism.
After $t_{\max}$ iterations, the algorithm outputs all $N$ hard bits.

\begin{algorithm}[htb]
	\caption{Parallel decoding framework.}
	\label{alg:stage-permute-decoder}
	\begin{algorithmic}[1]
		\REQUIRE ~~\\
		The received signal $\mathbf{y} = \{y_k, k=0\cdots N-1\}$;\\
		\ENSURE ~~\\ %??????Output
        The recovered codeword: $\hat{\mathbf{x}} = \{\hat{x_k}, k = 0 \cdots N-1\}$;\\
		\STATE Initilize: $L_{ch, k} \triangleq \frac{2y_k}{\sigma^2}$ for $k=0\cdots N-1$; \\
		\FOR {iterations: $t=1 \cdots t_{\max}$}
		\STATE Select decoding graph: $\Lambda = ~ (t\%2)~?~{\cal G}~:~{\cal G_\pi}$;
		\FOR{inner component decoders: $i=0,1 \cdots \sqrt{N}-1$ (in parallel)}		
		\STATE $L_{(i,j)}^t = Lgen(L_{ch, k}, \hat{c}_{(j,i)}^{t-1}, \hat{c}_{(i,j)}^{t-2}, e_j^{t-1}), \forall j$;
		\STATE $\hat{c}_i^{t},e_i^t =SubDecoder(L_i^t,\mathcal{F}_{\Lambda,i})$;
		\ENDFOR
		\ENDFOR
		\FOR{$\forall (i,j)$}
		\STATE  $\hat{x}_{k} = \hat{c}_{(i,j)}^{t_{max}}; (i,j)\leftrightharpoons k$ is described in \eqref{k_location}.
		\ENDFOR
	\end{algorithmic}
\end{algorithm}

\subsection{Component decoder: $SubDecoder()$}\label{sect:alg:subdecoder}
The component decoder $SubDecoder()$~\cite{SC_COMP} is described in Algorithm~\ref{alg:componentdecoder}.
Before each SC decoding, error detection is performed.
This can be achieved by applying a syndrome check based on hard decisions (line 2$\sim$5 in Algorithm~\ref{alg:componentdecoder}).
The cases with detected errors are denoted by Type-1 and otherwise Type-2.

The error detector brings two-fold advantages. On the one hand, since Type-$2$ component codes require no further SC decoding, this approach reduces power consumption. If all $\sqrt{N}$ sub-codes pass error detection, decoding can be early terminated for further power saving.
On the other hand, the error detection result provides us with additional information that the input LLRs of Type-2 component codes are more reliable than those of Type-1.
Such information can be used to improve the overall performance by estimating the input LLRs from the hard outputs.

\begin{algorithm}[htb]
	\caption{The $i$-th $SubDecoder()$}
	\label{alg:componentdecoder}
	\begin{algorithmic}[1] %??1 ??????????
		\REQUIRE ~~\\ %????????Input
		The input LLRs: $L_i^t$;\\
		Frozen set: $\mathcal{F}_{\Lambda,i}$.
		\ENSURE ~~\\ %??????Output
		Binary output $\hat{c}_i^t$, error detected indicator $e_i^t$;
        \STATE $e_i^t=False$;
		\STATE Hard decisions: \\
        \qquad $\hat{c}_{(i,j)}^{t} = L_{(i,j)}^t<0, j\in\{0,1,\dots,\sqrt{N}-1\}$;
		\STATE Vector $\hat{u}_{i}^{t} = \hat{c}_i^t\times G_{\sqrt{N}}$;
		\STATE Syndrome check:
		\IF {$\hat{u}_{(i,j)}^{t} \neq 0, \forall j\in \mathcal{F}_{\Lambda,i}$}
		\STATE $e_i^t=True$;
		\STATE SC decoding: $\hat{c}_i^{t}=SCDecoder(L_i^t,\mathcal{F}_{\Lambda,i})$; \\
		\ENDIF
	\end{algorithmic}
\end{algorithm}

\subsection{Input LLR generator: $Lgen()$}\label{sec:lcal}
In each iteration, the input LLRs are calculated by
\begin{itemize}
\item Type-1: $e_j^{t-1}=1$
\begin{equation}\label{equ:type-1}
L_{(i,j)}^t = L_{ch, k} + \frac{2\alpha^t}{\sigma^2}(1-2\hat{c}_{(j,i)}^{t-1})-\frac{2\beta^t}{\sigma^2}(1-2\hat{c}_{(i,j)}^{t-2}),
\end{equation}
\item Type-2: $e_j^{t-1}=0$
\begin{equation}\label{equ:type-2}
L_{(i,j)}^t = L_{ch, k} + \frac{2\gamma^t}{\sigma^2}(1-2\hat{c}_{(j,i)}^{t-1}),
\end{equation}
\end{itemize}
where a set of damping factors ($\alpha^t$, $\beta^t$, $\gamma^t$) are defined for each iteration, and $k$ is calculated from $(i,j)$ according to \eqref{k_location}.
The specific values of damping factors are optimized in \cite{SC_COMP}.

Since SC decoder is invariant to input LLR scaling, we can cancel noise variance $\sigma^2$ during LLR initialization. By multiplying both sides of  \eqref{equ:type-1} and \eqref{equ:type-2} by $\frac{\sigma^{2}}{2}$, the equations are simplified as follows
\begin{equation}\label{equ_new_LLR}
    \left\{
        \begin{array}{lr}
        \tilde{L}_{i,j}^t = y_{k} + \alpha^t(1-2\hat{c}_{(j,i)}^{t-1}) - \beta^t(1-2\hat{c}_{(i,j)}^{t-2}),\;e_j^{t-1} = 1,\\
        \tilde{L}_{i,j}^t = y_{k} + \gamma^t(1-2\hat{c}_{(j,i)}^{t-1}),\;e_j^{t-1} = 0,
        \end{array}
    \right.
\end{equation}
where $\tilde{L} \triangleq \frac{L\times{\sigma^{2}}}{2}$ and $y$ is the received signal.

We use a pair of new coefficients to replace $\alpha$ and $\beta$:
\begin{equation*}
    \left\{
        \begin{array}{lr}
        \delta^t = \alpha^t + \beta^t, \\
        \theta^t = \alpha^t - \beta^t.
        \end{array}
    \right.
\end{equation*}

Hence \eqref{equ_new_LLR} can be replace by \eqref{equ_new_LLR2}.
\begin{equation}\label{equ_new_LLR2}
    \tilde{L}_{i,j}^t = {y}_{k} + \Delta_{i,j}^{t}(1-2\hat{c}_{(j,i)}^{t-1}),
\end{equation}
in which $\Delta_{i,j}^{t}$ is determined by the binary $e_j^{t-1}$ and the hard outputs of the previous two iterations $c_{j,i}^{t-1},c_{i,j}^{t-2}$ as in \eqref{equ:delta}. The input signal calculation reduces to only one addition operation.

\begin{equation}\label{equ:delta}
	\Delta_{i,j}^{t}=\left\{
	\begin{array}{lcl}
    \delta^t, \text{where}\;e_j^{t-1}=1\;\text{and}\;c_{j,i}^{t-1} \neq c_{i,j}^{t-2},\\
	\theta^t, \text{where}\;e_j^{t-1}=1\;\text{and}\;c_{j,i}^{t-1} = c_{i,j}^{t-2},\\
    \gamma^t, \text{where}\;e_j^{t-1}=0.\\
	\end{array} \right.
\end{equation}

\section{An ASIC implementation}\label{section_implementations}
In this section, we present the ASIC implementation of a PDF-SC decoder in TSMC $16nm$ process for $N=16384$. The hardware optimization addresses both the SC decoders for component codes and the overall parallel decoding framework for $G_N$-coset codes.

\subsection{Bit Quantization}
Lower precision quantization is the key to higher throughput, thanks to its reduced implementation area and increased clock frequency.
As a tradeoff, performance loss is expected.
To maximize throughput while retaining performance, we must determine an appropriate quantization width.
Specifically, we use simulation to find out the smallest quantization width of a fixed-point decoder within $0.1dB$ performance loss from a floating decoder.

\begin{figure}[]
	\centering
	\includegraphics[width=3.5in]{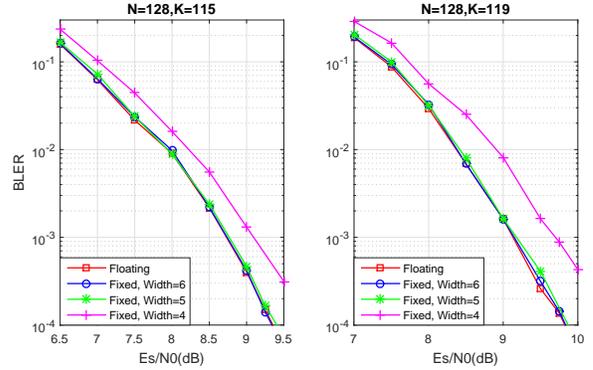}
	\caption{Sub-Decoder Performance Comparison between Floating Point and Fixed Point.}
	\label{sub-decoder-performance}
\end{figure}

First, we compare the performance of component codes under Algorithm~\ref{alg:componentdecoder}. We test two cases $N_{sub}=128,K_{sub}=115$ and $N_{sub}=128,K_{sub}=119$ with different quantization widths. According to Fig.\ref{sub-decoder-performance}, 6-bit quantization achieves the same performance as floating-point, 5-bit quantization incurs $<0.1dB$ loss, and 4-bits quantization yields significant loss. Therefore, we set the quantization width to 5 or 6 bits.

\begin{figure}[]
	\centering
	\includegraphics[width=3.5in]{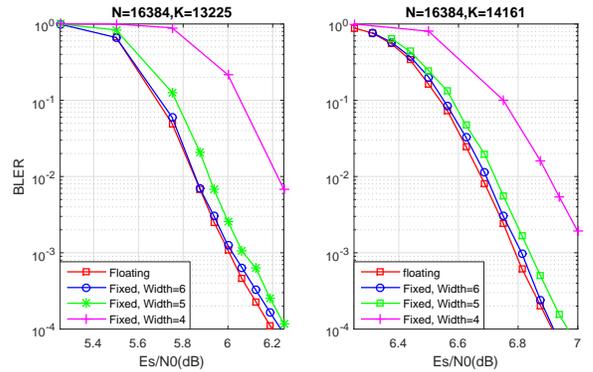}
	\caption{Decoder Performance Comparison between Floating Point and Fixed Point.}
	\label{decoder-performance}
\end{figure}

We then simulate the BLER performance of the long codes $N=128^2,K=115^2$ and $N=128^2,K=119^2$ under different quantizations, as shown in Fig.~\ref{decoder-performance}. Similarly, 6-bit quantization has no performance loss, and 5-bit quantization only incurs $<0.1dB$ loss for both code rates. Again, 4-bit quantization suffers from performance degradation.

\subsection{SC Core Optimization}
A component SC decoder is optimized via Rate-0 nodes, Rate-1 nodes\cite{SSC}, single parity check (SPC) nodes and repetition nodes (REP)\cite{FSSCL}. The decoder skips all Rate-0 nodes, parallelizes Rate-1, SPC and REP nodes for code blocks shorter than 32. If neither applies, maximum-likelihood (ML) multi-bit decision~\cite{FSSC} is employed for 4-bit blocks.

The architecture of SC decoders used here is described in \cite{Jiajie}, with supported code length reduced from 32768 to 128 to save area.
With TSMC $16nm$ technology, an SC core synthesizes to 4,100$\mu m^2$ area. Under 1.05$Ghz$ clock frequency, its latency is shown in Table~\ref{table_sclatency}.

\begin{table}
  \renewcommand{\arraystretch}{1.15}
    \caption{Sub-Decoder Decoding Latency for N=128 Polar codes}
    \label{table_sclatency}
    \centering
    \begin{tabular}{|c|c|c|c|c|c|}
    \hline
       \multicolumn{2}{|c|}{Information Bits(K)}  & 111  & 115    & 119  &122 \\
    \hline
      \multirow{2}*{Latency} &(Cycle)     & 24  & 19     & 18   &13  \\
    \cline{2-6}
       &(ns)       & 22.8  &18.05        &17.1    &12.35  \\
    \hline
    \end{tabular}
   %   \end{threeparttable}
\end{table}

\subsection{SC Core Sharing}
A unique design that significantly reduces area is called ``SC core sharing''.
In particular, we bind four SC cores as a sub-decoder group. The four SC cores share input/output pins, LLR updating circuits and error detector related components.
The sharing reduces a lot of local computation resources and global routing resources, but increases the latency between iterations. However, the overall area efficiency and power efficiency are improved.

\begin{figure}[]
	\centering
	\includegraphics[width=3.in]{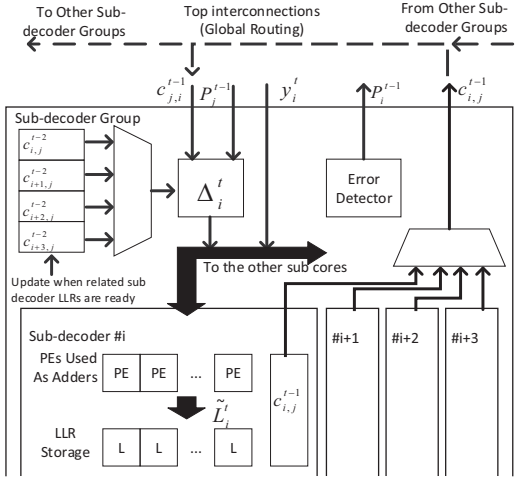}
	\caption{The Sub-Decoders Group architecture.}
	\label{sub-archi}
\end{figure}

In addition, we reuse the adders in the SC core to perform the input LLR addition in \eqref{equ_new_LLR2}. These adders were used for the $f_+$-function calculation in each processing element (PE)~\cite{Dec:LLR_based_SCL}. Altogether, $30\%$ area can be saved for each sub-decoder group. Fig.~\ref{sub-archi} shows the architecture of a sub-decoder group, including how adders in the PEs are reused. We run the synthesis flow with TSMC $16nm$ process, and the resulting area of a sub-decoder group is 19,400$\mu m^2$.

\subsection{Global Layout}
\begin{figure}[]
	\centering
	\includegraphics[width=3.1in]{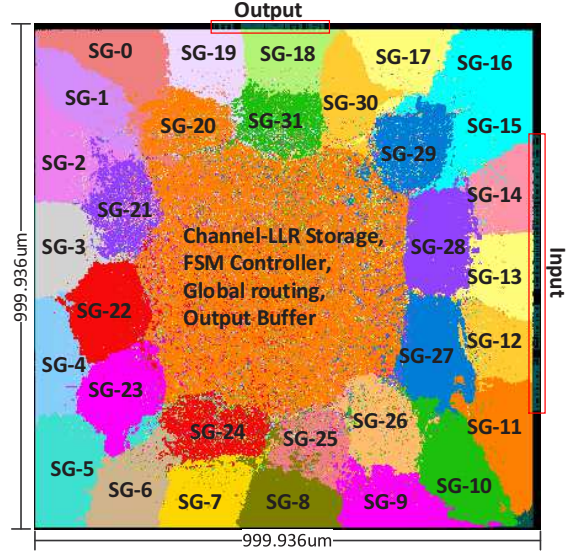}
  	\caption{The global layout of ASIC for parallel decoding.}
	\label{top-layout}
\end{figure}

Combining all algorithmic and hardware optimizations, the synthesized decoder ASIC requires $999.936\mu m\times999.936\mu m\approx 1.00mm^2$ area. The global layout is presented in Fig.~\ref{top-layout}. In the center of the layout is the top logic, including the input channel LLR storage, finite state machine (FSM) controller, interleaved connection routing and output buffer. The aforementioned 32 sub-decoder groups (SG in the figure) are placed around the top logic, highlighted by different colors.

\section{Key Performance Indicators}
\begin{table}
  \renewcommand{\arraystretch}{1.15}
    \caption{Decoder Area Efficiency}
    \label{table_throughput}
    \centering
    \begin{tabular}{|c|c|c|c|c|c|c|}
    \hline
      Info & Iter- & Es/N$_0$ & Latency & Area Eff. & \multicolumn{2}{c|}{Convert to}\\
      \cline{6-7}
      size &ation &($dB$) & (ns) &(Gbps/$mm^2$) &$10nm$ &$7nm$ \\
    \hline
      \multirow{5}*{\tabincell{c}{13225}} & 4  & 7.14  &91.2  &135.07	&310.66	&596.47 \\
    \cline{2-7}
       & 5  & 6.82  &114   &108.06	&248.53	&477.17 \\
    \cline{2-7}
       & 6  & 6.55  &136.8 &90.05	&207.11	&397.64 \\
    \cline{2-7}
       & 7  & 6.36  &159.6 &77.18	&177.52	&340.84 \\
    \cline{2-7}
       & 8  & 6.20  &182.4 &67.53	&155.33	&298.23  \\
    \hline
    \multirow{5}*{\tabincell{c}{14161}}  & 4  & 7.79  &87.4	&150.92	&347.11	&666.45 \\
    \cline{2-7}
       & 5  & 7.48  &109.25	&120.73	&277.69	&533.16 \\
    \cline{2-7}
       & 6  & 7.22  &131.1	&100.61	&231.41	&444.30 \\
    \cline{2-7}
       & 7  & 7.06  &152.95	&86.24	&198.35	&380.83 \\
    \cline{2-7}
       & 8  & 6.97  &174.8	&75.46	&173.55	&322.22 \\
    \hline
    \end{tabular}
   %   \end{threeparttable}
\end{table}

\begin{table*}
  \renewcommand{\arraystretch}{1.05}
    \caption{COMPARISON WITH FABRICATED ASIC OF TRADITIONAL POLAR DECODER}
    \label{table_comparsion}
    \centering
    \begin{tabular}{lcccccccc}
      Implementation    & This Work & This Work & This Work & This Work & \cite{Jiajie} & \cite{Jiajie} & \cite{Jiajie} & \cite{Jiajie}\\
      \hline
      Construction      &$G_N$-coset&$G_N$-coset&$G_N$-coset&$G_N$-coset& Polar       & Polar	     & Polar 	     &Polar \\
      Decoding Algorithm         & PDF-SC       & PDF-SC       & PDF-SC       & PDF-SC       &SC             &SC	     &CA-SC-List	     &CA-SC-List \\
      List size / Iterations  & 5         & 8         & 5         & 8         & 1             & 1              & 8             & 8    \\
      Code Length       & 16384     & 16384     & 16384     & 16384     &32768          &32768   	     &16384	         &16384 \\
      Code Rate         & 0.807     & 0.807     & 0.864     & 0.864     &0.807          &0.864	         &0.807	         &0.864 \\
      EsN0@BLER=$10^{-4}$ &6.82     & 6.20      & 7.48      & 6.97      &5.61           &6.49            &5.24           &6.13 \\
      \hline
      Technology        &\multicolumn{8}{c}{All in TSMC $16nm$} \\
      Clock Ferquency($Ghz$)&1.05     &1.05     &1.05       &1.05       &1.00       &1.00       &1.00       &1.00 \\
      Throughtput ($Gbps$)      & 108.06    & 67.53     &120.73     &75.46      &4.16 	    &4.56       &0.91       &1.01 \\
      Area($mm^2$)      & 1.00      & 1.00      & 1.00      &1.00       &0.35       &0.35       &0.45	    &0.45 \\
      Area Eff($Gbps/mm^2$) & 108.06  & 67.53   &120.73     &75.46      &11.89   	&13.02      &2.02       &2.24 \\
      \hline
    \end{tabular}
   %   \end{threeparttable}
\end{table*}

\begin{figure*}[]
	\centering
	\includegraphics[width=6.5in]{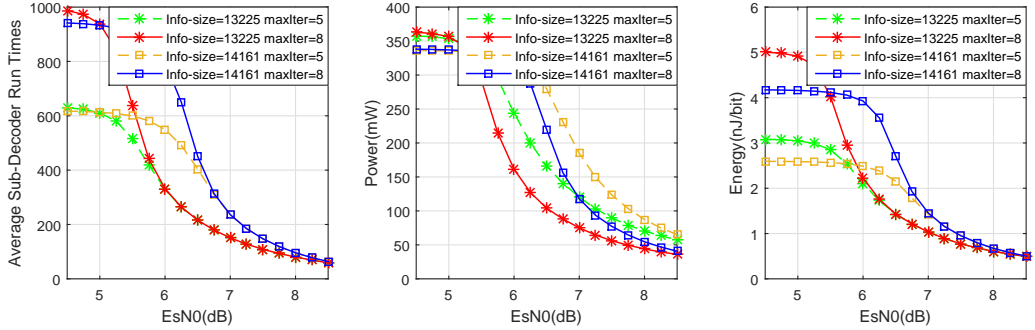}
  	\caption{Left: The Sub-decoder running times during per packet decoding. Center: Decoding Power efficiency (mW/mm$^2$). Right: Energy efficiency (pJ/bit).}
	\label{rt-pw-eg}
\end{figure*}

The key performance indicators (KPIs) are examined. First of all, we evaluate the area efficiency using equation $Area\,Eff(Gbps/mm^{2}) = \frac{Info\,Size(bits)}{Latency(ns)\times Area(mm^{2})}$\footnote{The error detector can early terminate the decoding, but its worst-case latency is guaranteed by maximum iteration times, as required by most practical communication systems.}. The proposed decoder can reach up to hundreds of gigabits per second within one square millimeter.
The evaluated throughput in given in Table~\ref{table_throughput}\footnote{The third column ``Es/N0'' is chosen such that BLER$=10^{-4}$.}. With TSMC $16nm$ process, the area efficiency for code rate $13225/16384$ and $14161/16384$ are $67Gbps/mm^2$ and $75Gbps/mm^2$ when the maximum number of iterations is eight.
If we reduce to five iterations by allowing $0.5db$ performance loss, the area efficiency can reach $108Gbps/mm^2$ and $120Gbps/mm^2$.
The estimated throughput under $10nm$ and $7nm$ technologies can be converted from $16nm$\footnote{The converting ratios including cell density ratio and speed improvement ratio are obtained from the TSMC process introductions~\cite{TSMC10nm,TSMC7nm}.}. With the more advanced $7nm$ process, the throughput is as high as $477Gbps/mm^2$ and $533Gbps/mm^2$, which are much higher than the $100Gbps/mm^2@7nm$ target given in the EPIC project~\cite{EPIC}. Note that the KPI is achieved at code length $16384$, which exhibits significant coding gain over codes with length $1024$~\cite{SC_COMP}. With future technologies of $5nm$ and below, it is promising to achieve an extremely challenging target of $1Tbps/mm^2$.

The area efficiency is also compared with a highly-optimized and fabricated ASIC polar\footnote{The polar codes are constructed by Gaussian approximation at Es/N0=$5.3dB$, $6.3dB$, $5.3dB$ and $6.0dB$ for ($N,R$)=($16384,0.807$), ($16384,0.864$), ($32768,0.807$) and ($32768,0.864$) respectively. $N$ is code length and $R$ is code rate.} decoder in~\cite{Jiajie}. For both code rates, the throughput of the proposed decoder  (with five iterations) is nine times as fast as a polar fast-SC decoder, and 53 times that of a CA-SC-List-8 decoder\footnote{In~\cite{Jiajie}, the fabricated ASIC SC decoder supports code length $32768$.}. Detailed comparison results can be found in Table~\ref{table_comparsion}.

We further evaluate the average running time and power consumption (per packet).
In the lower SNR region, longer decoding time and higher power consumption are observed.
But in the higher SNR region, both decoding time and power consumption are much smaller, thanks to the built-in error detection and early termination function described in section~\ref{sect:alg:subdecoder}.
Specifically, only 15\% component codes cannot pass the error detection while the rest 85\% can skip SC decoding.
The power consumption (mW/mm$^2$) and energy efficiency (pJ/bit) follow similar trend to the average running time.
When Es/N0$>8dB$, the power consumption is smaller than $100mW/mm^2$. The energy efficiency is around $1pJ/bit$, again meeting the target proposed in~\cite{EPIC}.
These results\footnote{Note that the power consumption with eight iterations is lower than that with five iterations. This is due to the lower error detection successful rate during the first five (1-5) iterations. The last three (5-8) iterations consume much lower power, which reduced the averaged power level.} are plotted in Fig.~\ref{rt-pw-eg}. The power consumption and energy efficiency are evaluated with TSMC $16nm$ process.

\section{Conclusions}
\label{section_conclusions}
In this paper, we present an ASIC implementation of high-throughput $G_{N}$-coset codes. The parallel decoding framework leads to a hardware with high area efficiency and low decoding power consumption. An area efficiency of $120Gbps/mm^{2}$ is achieved within approximately $1mm^{2}@16nm$ process. The power consumption is as low as $100mW$ and energy efficiency is around $1pJ/bit$. Scaled to $7nm$ process, the area efficiency can reach $533Gbps/mm^{2}$. It confirms that $G_N$-coset codes can meet the high-throughput demand in next-generation wireless communication systems.
\bibliographystyle{IEEEtran}

\end{document}